\documentclass[twocolumn,showpacs,prl,amsmath,amssymb]{revtex4}

\usepackage{graphicx}
\usepackage{dcolumn}
\usepackage{bm}

\begin{document}

\preprint{APS/123-QED}

\title{Electronic structure reconstruction by orbital symmetry breaking in IrTe$_2$}

\author{D.~Ootsuki$^1$}
\author{S.~Pyon$^2$}
\author{K.~Kudo$^2$}
\author{M.~Nohara$^2$}
\author{M.~Horio$^3$}
\author{T.~Yoshida$^3$}
\author{A.~Fujimori$^3$}
\author{M.~Arita$^4$}
\author{H.~Anzai$^4$}
\author{H.~Namatame$^4$}
\author{M.~Taniguchi$^{4,5}$}
\author{N.~L.~Saini$^{6,1}$}
\author{T.~Mizokawa$^{1}$}

\affiliation{$^1$Department of Complexity Science and Engineering \& 
Department of Physics, University of Tokyo, 5-1-5 Kashiwanoha, Chiba 277-8561, Japan}
\affiliation{$^2$Department of Physics, Okayama University, Kita-ku, Okayama 700-8530, Japan}
\affiliation{$^3$Department of Physics, University of Tokyo, 7-3-1 Hongo, Tokyo 113-0033, Japan}
\affiliation{$^4$Hiroshima Synchrotron Radiation Center, Hiroshima University, Higashi-hiroshima 739-0046, Japan}
\affiliation{$^5$Graduate School of Science, Hiroshima University, Higashi-hiroshima 739-8526, Japan}
\affiliation{$^6$Department of Physics, University of Roma "La Sapienza" Piazzale Aldo Moro 2, 00185 Roma, Italy}

\date{\today}

\begin{abstract}
We report an angle-resolved photoemission spectroscopy (ARPES) study 
on IrTe$_2$ which exhibits an interesting lattice distortion below 270 K 
and becomes triangular lattice superconductors by suppressing 
the distortion via chemical substitution or intercalation. 
ARPES results at 300 K show multi-band Fermi surfaces with six-fold symmetry 
which are basically consistent with band structure calculations. 
At 20 K in the distorted phase, topology of the inner Fermi surfaces is 
strongly modified by the lattice distortion. 
The Fermi surface reconstruction by the distortion depends on the orbital 
character of the Fermi surfaces, suggesting importance of Ir 5$d$ and/or 
Te 5$p$ orbital symmetry breaking.
\end{abstract}

\maketitle

Transition-metal compounds with multi-band Fermi surfaces 
often exhibit rich and interesting physical properties such as 
spin-charge-orbital order and superconductivity which originate 
from the topology of their multi-band Fermi surfaces.
For example, the multi-orbital electronic structures of transition-metal oxides 
and chalcogenides including CuIr$_2$S$_4$ and Ca$_{2-x}$Sr$_x$RuO$_4$ provide 
various metal-insulator transitions with spin-charge-orbital ordering
\cite{Imada1998, Nagata1994, Radaelli2002, Nakatsuji2000}.
Also the multi-band structure of the Fe $3d$ orbitals play important 
roles in superconductivity and magnetism of Fe pnictides and 
chalcogenides such as LaFeAsO$_{1-x}$F$_x$ \cite{Kamihara2008}.
Recently, Pyon {\it et al.} \cite{Pyon2012} and Yang {\it et al.} 
\cite{Yang2012} have discovered interesting interplay between 
lattice distortion and superconductivity in triangular lattice 
IrTe$_2$ in which multi-band Fermi surfaces are expected to 
play significant roles. Since the large spin-orbit interaction 
of Ir 5$d$ electrons is expected to entangle the spin and orbital
degrees of freedom in IrTe$_2$ and the derived superconductors,
Yang {\it et al.} pointed out that the IrTe$_2$ system provides 
a new playground to explore and/or realize topological quantum states, 
which are currently attracting great interest in physics community \cite{Yang2012}.

IrTe$_2$ exhibits a structural phase transition at $\sim$ 270 K 
from the trigonal (P3m-1) to the monoclinic (C2/m) structure
accompanied by anomalies of electrical resistivity and 
magnetic susceptibility \cite{Matsumoto1999}. 
When the lattice distortion is suppressed by chemical substitution 
of Pt or Pd for Ir or intercalation of Pd, IrTe$_2$ becomes 
superconductors \cite{Pyon2012,Yang2012}.
An electron diffraction study by Yang {\it et al.} \cite{Yang2012} observed 
the superlattice peaks with wave vector of $q$ = (1/5, 0, -1/5) below the 
structural transition temperature. Such superstructure can be explained 
by charge density wave (CDW) driven by perfect or partial nesting of multi-band 
Fermi surfaces.
In multi-band Fermi surfaces derived from Ir 5$d$ and Te 5$p$ orbitals, 
the nesting character can be enhanced by orbitally-induced Peierls 
mechanism \cite{Khomskii2005}.  
In addition, charge modulation of Ir 5$d$ electrons is indicated
by an Ir 4$f$ x-ray photoemission study \cite{Ootsuki2012}.
On the other hand, a recent optical study by Fang {\it et al.}
on single crystal samples shows that there is no gap opening 
expected for CDW and, instead, band structure is reconstructed
over a broad energy scale up to $\sim$ 2 eV \cite{Fang2012}. 
Fang {\it et al.} conclude that the structural transition 
of IrTe$_2$ is not of CDW type but of a novel type driven 
by Te 5$p$ holes \cite{Fang2012}.

In this context, it is very interesting and important to 
study the geometry of multi-band Fermi surfaces of IrTe$_2$
using angle-resolved photoemission spectroscopy (ARPES). 
In the present ARPES study, above the transition temperature,
the flower-shaped outer Fermi surface and the inner Fermi surfaces 
like six connected beads, which are predicted by band structure 
calculations, are partly identified. 
Across the structural transition, the topology of the inner Fermi 
surfaces is modified more strongly than that of the outer Fermi surface.
Below the transition temperature, the inner Fermi surfaces
consist of two straight portions, suggesting Fermi surface nesting.
However, clear gap opening expected for CDW is not observed
in the ARPES spectra, consistent with the optical study \cite{Fang2012}.
Instead, spectral weight is partially suppressed at specific points of 
the straight Fermi surfaces.


Single crystal samples of IrTe$_2$ were prepared using 
a self-flux method \cite{Fang2012,Pyon2012b}.
The ARPES measurements were carried out at beamline 9A, 
Hiroshima Synchrotron Radiation Center using a SCIENTA R4000 analyzer
with circularly polarized light of photon energy $h\nu$ = 23 eV.
The data were collected at 300 K and 20 K with an angular resolutions of 
$\sim$ 0.3$^{\circ}$ and energy resolution of 18 meV for excitation energy 
of $h\nu$ = 23 eV.
The incident beam is 50$^{\circ}$ off the sample surface. 
The base pressure of the spectrometer was in the $10^{-9}$ Pa range. 
The samples were cleaved at $300$ K under the ultrahigh vacuum and 
cooled across the structural transition, and then warmed 
to 300 K to check the reproducibility at 300 K.
The samples were oriented by {\it ex situ} Laue measurements.
The spectra were acquired within 8 hours after the cleavage.
Binding energies were calibrated using the Fermi edge of
gold reference samples.

\begin{figure}
\includegraphics[width=8cm]{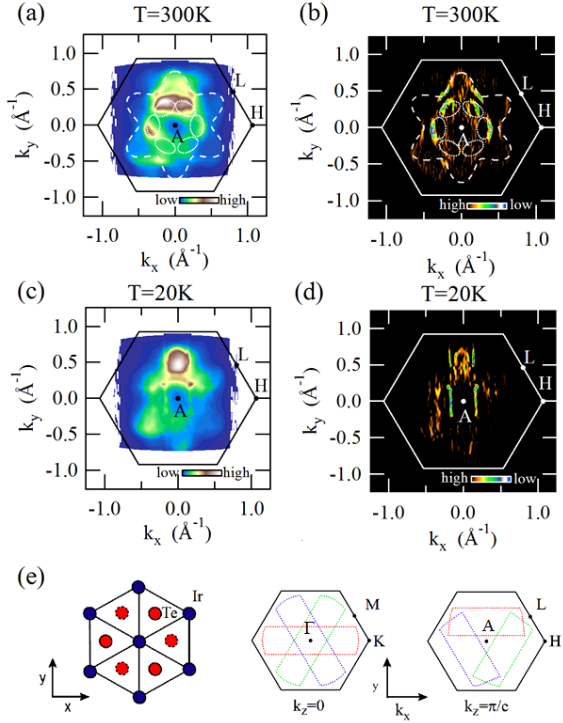}
\caption{
(color online) 
(a) Fermi surface map and (b) its second derivative map 
of IrTe$_2$ for $h\nu$ = 23 eV taken at 300 K.
(c) Fermi surface map and (d) its second derivative map 
of IrTe$_2$ for $h\nu$ = 23 eV taken at 20 K.
The integration energy window of  $\pm$5 meV at the Fermi level ($E_F$).
The center of the hexagon roughly corresponds to the A point for 
$h\nu$ = 23 eV. For 300 K, the flower-shaped outer Fermi surface
and the inner Fermi surfaces with six-fold symmetry 
are schematically shown by the dashed and dotted curves, respectively.
(e) Schematic drawings for the Ir triangular lattice and
the hexagonal Brillouin zone at $k_z$ = 0 and $k_z$ = $\pi/c$.
The Te ions indicated by solid (dotted) circles are
located above (below) the Ir plane. 
The thin solid curves indicate possible Brillouin zone boundaries 
for possible three domains considering the superstructure reported in ref. 7.
}
\end{figure}

\begin{figure}
\includegraphics[width=8cm]{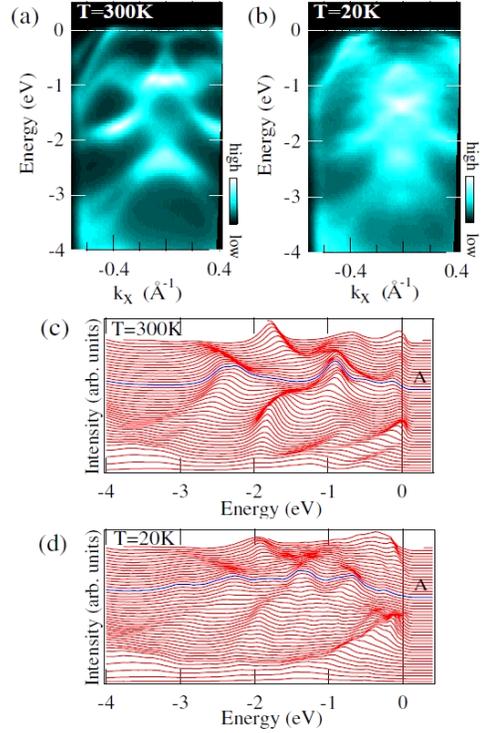}
\caption{
(color online) 
Broad-range band dispersions along the A-H direction of IrTe$_2$ for $h\nu$ = 23 eV 
taken at 300 K (a) and at 20 K (b). Broad-range energy distribution curves 
along the A-H direction of IrTe$_2$ for $h\nu$ = 23 eV taken at 300 K (c) 
and at 20 K (d).
}
\end{figure}

\begin{figure}
\includegraphics[width=8cm]{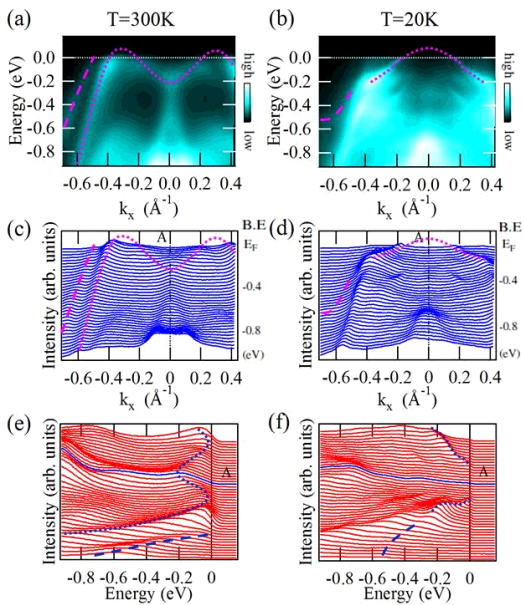}
\caption{
(color online) 
Near-$E_F$ band dispersions along the A-H direction 
of IrTe$_2$ for $h\nu$ = 23 eV taken at 300 K (a)
and at 20 K (b). Near-$E_F$ momentum distribution curves along
the A-H direction of IrTe$_2$ for $h\nu$ = 23 eV taken 
at 300 K (c) and at 20 K (d). Near-$E_F$ energy distribution curves
along the A-H direction of IrTe$_2$ for $h\nu$ = 23 eV taken at
300 K (e) and at 20 K (f). The outer hole bands and the inner hole-like
bands are indicated by the dashed and dotted curves, respectively.
}
\end{figure}

\begin{figure}
\includegraphics[width=8cm]{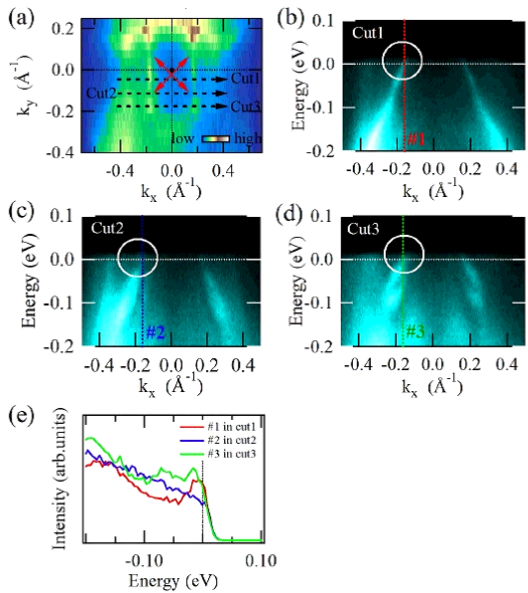}
\caption{
(color online) 
(a)Fermi surface mapping near the A point 
at 20 K for $h\nu$ = 23 eV.
(b-d)Near-$E_F$ band dispersions along the cuts parallel 
to the A-H direction at 20 K for $h\nu$ = 23 eV.
(e) Near-$E_F$ energy distribution curves at the 
selected Fermi surface points at 20 K.
}
\end{figure}

The Fermi surface mapping of IrTe$_2$ measured at 300 K above 
the structural transition temperature are displayed in Figure 1(a).
At $h\nu$ = 23 eV, the momentum perpendicular to the Ir plane approximately 
corresponds to $\pi/c$, where $c$ is the out-of-plane lattice constant, 
and the center of the hexagonal Brillouin zone is the A point.
The direction from the A point to the L (H) point corresponds 
to the direction of Ir-Ir (Ir-Te) bond.
In Figs. 1(a), several Fermi surfaces can be identified 
as predicted by the band structure calculations \cite{Yang2012,Fang2012}
although the strong intensity asymmetry due to transition-matrix element 
effect does not allow perfect identification. In order to extract 
the shapes of the Fermi surfaces,
the second derivative along the cut direction $d^2I(k_x,k_y)/d^2k_x$ 
is plotted in Fig. 1(b).
The flower shape for the outer Fermi surface is more clearly
seen which is schematically indicated by the dashed curve.
In addition, the inner Fermi surfaces like six connected beads 
can be identified as indicated by the dotted curves although
effect of thermal excitations at 300 K tends to obscure 
the relatively small Fermi pockets.
The inner Fermi surfaces observed around the A point at 300 K
are roughly consistent with the prediction of band-structure 
calculations \cite{Yang2012,Fang2012}.

Figure 1(c) shows the Fermi surface mapping at 20 K well below the transition 
temperature.
Across the structural transition, in the region where the outer 
Fermi surface is close to the inner Fermi surfaces, while 
the outer Fermi surface at 300 K is observed separately
from the inner Fermi surfaces, the outer Fermi surface at 20 K 
disappears due to  partial gap opening or overlaps 
with the inner Fermi surfaces.
In addition, in going from 300 K to 20 K, intensity evolves 
in the six leaves of the outer Fermi surface flower
probably due to band folding with $q$ = (1/5, 0, -1/5).
In contrast to the limited effect on the outer Fermi surface, 
the inner Fermi surfaces dramatically change their shapes
by the structural transition.
The band structure calculations show that the outer Fermi surface 
is mainly constructed from the Ir 5$d$ $a_{1g}$ 
[$\frac{1}{\sqrt{3}}(XY + YZ + ZX)$] and Te 5$p_z$ orbitals, 
and that the inner Fermi surfaces mainly have the Ir 5$d$ $e_{g}'$ 
[$\frac{1}{\sqrt{3}}(XY+e^{\pm2\pi i/3}YZ+e^{\pm4\pi i/3}ZX)$] 
and Te 5$p_{x,y}$ orbital components (Here, the X-, Y-, and Z-axes are along 
the three Ir-Te bonds of a regular IrTe$_6$ octahedron). 
This assignment is supported by the good agreement between 
the ARPES result at 300 K and the band structure calculations \cite{Yang2012,Fang2012}.
As for the band structure change across the transition, 
the experimental result that the inner Fermi surfaces are more 
strongly affected by the transition is consistent with the band structure 
calculation by Fang {\it et al} \cite{Fang2012}. 
However, the geometry of inner Fermi surfaces at 20 K 
deviates from the prediction of the calculation.
Interestingly, two straight portions of Fermi surfaces are observed at 20 K. 
The straight portions are perpendicular to the A-H direction or 
the direction of Ir-Te bond. Therefore, both of the Te 5$p$ and Ir 5$d$ orbitals
would be involved in the structural transition if the straight Fermi surfaces 
are driven by orbitally-induced Peierls mechanism \cite{Khomskii2005}.

Figures 2(a) and (b) show the broad-range band dispersions along the A-H direction
at 300 K and 20 K, respectively. Whereas the broad-range band dispersions at 300 K  
roughly agree with the predictions of the band-structure calculations, 
those at 20 K deviate from the predictions \cite{Yang2012,Fang2012}. 
In going from 300 K to 20 K, broad-range band structures up to -3 eV are 
strongly modified, which is consistent with the optical study \cite{Fang2012}.
The broad-range spectral change is more clearly seen in the energy distribution
curves for 300 K and 20 K shown in Figs. 2(c) and (d), respectively. 
In going from 300 K to 20 K, spectral peaks at 300 K tend to be split into 
several structures probably due to complicated Jahn-Teller-like effect and
band folding effect due to the charge and orbital ordering with the (1/5, 0, -1/5)
superstructure.
Consequently, in the energy range from -0.2 eV to -3 eV, 
the spectral peaks at 20 K are much broader than those at 300 K.

On the other hand, spectral peaks from $E_F$ to -0.2 eV are rather sharp at 20 K 
compared to those at 300 K as shown in Fig. 3. For 300 K, the outer hole band
is indicated by the dashed curve in Fig. 3(a) which form the flower-shaped outer 
Fermi surface of Fig. 1(a). The inner hole-like band indicated by the dotted curves
in Fig. 3(a) creates the hole pockets which corresponds to the inner Fermi surfaces 
of Fig. 1(a). By comparing between Figs. 3(a) and (b),
while the outer band at 300 K is observed separately from the inner one,
the outer band at 20 K disappears near $E_F$. 
Across the transition, the outer band is shifted towards the inner one 
in this momentum region and is probably gapped due to the interaction 
with the inner band. This is consistent with the partial disappearance
of the outer Fermi surface in Fig. 1(c).
The inner hole-like band at 300 K is also strongly affected by the structural transition.
The band located around $\sim$ -0.15 eV of the A point at 300 K disappears at 20 K probably 
because it is shifted above $E_F$. Consequently, the hole band indicated 
by the dotted curve in Fig. 3(b) crosses $E_F$ at 20 K and form the straight portions 
of the Fermi surfaces of Fig. 1(b). Such band reconstruction cannot be explained
by a simple band folding picture, indicating orbital reconstruction by 
Jahn-Teller-like effect.

The Fermi surface mapping around the A point at 20 K is shown in Fig. 4(a).
In general, straight Fermi surfaces with nesting wave vector $q$ 
are expected to be gapped due to density wave formation with $q$. 
In IrTe$_2$, instead of gap opening, spectral weight at $E_F$ is 
partially suppressed at specific points of the straight Fermi surfaces. 
In cuts 1 and 3 along the A-H direction [Figs. 4(b) and (d)],
the hole band clearly crosses $E_F$ and the spectral weight 
at $E_F$ is not suppressed. On the other hand, in cut 2 [Fig. 4(c)],
the intensity of the hole band is suppressed near $E_F$ as seen
in the EDC plot of Fig. 4(e). 
There are four points where the spectral weight at $E_F$ is suppressed
as seen in Fig. 4(a).
Such spectral weight suppression at the specific points (cold spots) would be 
related to the origin of the superstructure of bulk IrTe2 since the wave vectors 
connecting the two cold spots [indicated by the arrows in Fig. 4(a)] are 
approximately 2/5 of the A-L distance or 1/5 of the L-L' distance, partly 
consistent with its period. However, the partial spectral weight suppression 
would be due to surface effect or transition-matrix element effect, and 
no decisive conclusion can be obtained at the present stage.
Here, it should be noted that the observed Fermi surfaces correspond 
to one of the Brillouin zone boundaries for possible domains. 
However , the crystal structure of low temperature phase is highly 
controversial (refs. 7, 11, and 13), and that, at the present stage, 
it is difficult to discuss relationship between the observed Fermi surfaces 
and the band folding due to the superstructure.

In conclusion, above the transition temperature, the observed Fermi 
surfaces and band dispersions are consistent with the band structure 
calculations. The flower-shaped outer Fermi surface (hole character)
with six-fold symmetry and the inner Fermi surfaces (hole pockets) 
are observed. 
Across the structural transition, the geometry of 
the inner Fermi surfaces is strongly modified. 
In the distorted phase, the inner Fermi surfaces
consist of two straight portions, suggesting that nesting character
is enhanced. However, the gap opening expected for CDW is not observed
in the ARPES spectra, consistent with the optical study.
Also the electronic structure up to $\sim$ -3 eV is reconstructed
by the lattice distortion, which is also consistent with the optical
study. In addition, the spectral weight at $E_F$ is suppressed 
at the specific points of the straight Fermi surfaces, which would be 
related to the origin of the superstructure.

The authors would like to thank valuable discussions with D. I. Khomskii and H. Takagi.
This work was partially supported by a Grants-in-Aid for Young Scientists (B) (23740274, 24740238) 
from the Japan Society of the Promotion of Science (JSPS) and the Funding Program for 
World-Leading Innovative R\&D on Science and Technology (FIRST Program) from JSPS. 
The synchrotron radiation experiment was performed with the approval of 
HSRC (Proposal No.12-A-12).

\end{document}